\documentclass[a4paper,12pt,hidelinks]{article}
\usepackage[left=2.5cm,bottom=3cm,right=2.5cm,top=3cm]{geometry} 
\usepackage{lmodern}
\pdfoutput=1
\usepackage{graphicx}
\usepackage{xcolor}
\usepackage{amsmath}
\usepackage{amsfonts}
\usepackage{amssymb}
\usepackage{mathtools}
\usepackage{cite}
\usepackage[pagebackref=true]{hyperref}
\usepackage{tikz} 
\usepackage{calrsfs}
\usepackage{cleveref}
\crefname{figure}{Figure}{Figures}
\usetikzlibrary{decorations.markings,decorations.pathmorphing}
\usetikzlibrary{intersections}
\usetikzlibrary{shapes.misc}
\tikzset{cross/.style={cross out, draw=black, minimum size=2*(#1-\pgflinewidth), inner sep=0pt, outer sep=0pt},
cross/.default={1pt}}
\tikzstyle{singularity}=[red!50!black,line width=0.6,decorate,
                         decoration={zigzag,amplitude=2,segment length=6.17}]
\usepackage{graphicx} 
\usepackage{caption}
\usepackage{subcaption}
\graphicspath{{figures/}} 
\DeclareMathAlphabet{\mathcal}{OMS}{cmsy}{m}{n}
\usepackage{tikz} 
\usepackage{calrsfs}
\usepackage{cancel}
\usetikzlibrary{decorations.markings,decorations.pathmorphing}
\usetikzlibrary{intersections}
\usetikzlibrary{shapes.misc}
\tikzset{cross/.style={cross out, draw=black, minimum size=2*(#1-\pgflinewidth), inner sep=0pt, outer sep=0pt},
cross/.default={1pt}}
\tikzstyle{singularity}=[red!50!black,line width=0.6,decorate,
                         decoration={zigzag,amplitude=2,segment length=6.17}]

\usetikzlibrary{arrows.meta}

\tikzset{
flow/.style={
    thick,
    postaction={decorate},
    decoration={
        markings,
        mark=at position 0.55 with {\arrow{Stealth}}
    }
}
}
                         
\colorlet{myred}{red!70!black}
\colorlet{mygreen}{green!70!black}
\colorlet{mydarkblue}{blue!50!black}
\colorlet{myblue}{blue!13!white!90!black}
\colorlet{myblue2}{blue!45!black!40!}
\colorlet{myred}{red!70!black}
\colorlet{mygreen}{green!70!black}
\colorlet{mydarkblue}{blue!50!black}
\colorlet{myblue}{red!13!white!90!black}
\colorlet{myblue2}{red!30!white!70!black}
\colorlet{mypink}{gray!40!black}
\definecolor{ExecusharesWhite}{RGB}{255,255,243}
\usepackage{tcolorbox}
\usepackage{xpatch}

\newcommand{\doilink}[1]{%
  \href{https://doi.org/#1}{doi:#1}%
}

\newcommand{\arxivlink}[2]{%
  \href{https://arxiv.org/abs/#1}{arXiv:#1 [#2]}%
}

\newcommand{\inspirelink}[1]{%
  \href{https://inspirehep.net/literature/#1}{INSPIRE}%
}

\begin{document}
%
%
\begin{titlepage}
	\begin{flushright}
		IFT-UAM/CSIC-26-90
	\end{flushright}
	\vspace{.3in}
	\vspace{1cm}
	\begin{center}
		{\Large\bf\color{black} AdS Black Holes Are Short-Lived inside the Spectral Form Factor\\}
		\bigskip\color{black}
		\vspace{1cm}{
			{\large J.~L.~F. Barb\'on$^a$ and E. Velasco-Aja$^{a,b}$}
			\vspace{0.3cm}
		} \\[7mm]
		$^a$	{\it {Instituto de F\'{i}sica Te\'orica IFT-UAM/CSIC, Universidad Aut\'onoma de Madrid, Cantoblanco, 28049, Madrid, Spain}}\\[10pt]
		$^b$	{\it {Departamento de F\' \i sica Te\'orica, Universidad Aut\'onoma de Madrid, Cantoblanco 28049 Madrid, Spain}}\\[10pt]
		{\it E-mail:} \href{mailto:jose.barbon@csic.es}{\nolinkurl{jose.barbon@csic.es}}, \href{mailto:eduardo.velasco@uam.es}{\nolinkurl{eduardo.velasco@uam.es}}
	\end{center}
	\bigskip

\vspace{1cm}

\begin{abstract}
We analyze the contribution of AdS black hole saddles to the spectral form factor of $d$-dimensional CFTs with a gravity dual. At high temperatures, the analytically continued gravitational action of the large AdS black hole is expected to give a leading large-$N$ approximation to the `slope' of the spectral form factor. When the CFT is put on a $(d-1)$-dimensional spatial sphere of radius $\ell$, we show that such an evaluation implies unphysical behavior of $Z(\beta +it)$ at large $t\gg \ell$ for $d\equiv5 \,({\rm mod}\; 4)$. A Picard--Lefschetz analysis of a minisuperspace approximation to $Z(\beta+it) $ does reveal the required Stokes phenomena that restore consistency. 
It is found that for $d>3$, the large AdS black hole saddle loses dominance and subsequently disconnects from the relevant integration cycle at $t=O(\beta)$. For $d=3$, dominance is lost at a time of order $\beta$, while the Stokes disconnection takes place at $t=O(\ell)$.
After the large black hole saddle disconnects from the path-integral contour, the slope of the spectral form factor becomes dominated by the low-energy end of the spectrum. 
\end{abstract}
\end{titlepage}

      \tableofcontents

\section{Introduction}
Euclidean Gravitational Path Integrals are notoriously difficult to define from first principles, and yet they find in the thermodynamics of black holes an arena of spectacular success \cite{Hawrev, GHP, GH, HP}. This success has been greatly enhanced by the prominent role of these methods in our current understanding of holography. In particular, the semiclassical evaluation  of the gravitational path integral on Euclidean black hole saddles gives the large-$N$, strong coupling approximation to the canonical partition function of a CFT with a gravity dual,
\begin{equation}\label{sp}
Z(\beta) \approx  e^{-I(X^+_\beta)}\;,
\end{equation} 
where $I(X^+_\beta)$ is the Euclidean gravitational action evaluated on the Euclidean section of the large AdS$_{d+1}$ black hole at inverse temperature $\beta$, i.e., the `cigar' with AdS asymptotics \cite{W1, W2}. In the high-temperature domain, the resulting CFT free-energy density is proportional to $-N_* \,\beta^{-d}$, with $N_*$ the effective `central charge' of the CFT, related to the bulk parameters by $N_* \propto \ell^{d-1} /G$, where $\ell$ is the radius of curvature of AdS$_{d+1}$ and $G$ is Newton's constant. The precise proportionality constant is a prediction of the AdS/CFT correspondence. 

It is natural to expect that the gravitational evaluation of the partition function will continue to give the leading large-$N$ approximation when one slightly complexifies the temperature and computes $Z(\beta +it)$ for small $t$. This analytically continued partition function is the building block of the so-called Spectral Form Factor (SFF), $ \zeta (\beta, t) = |Z(\beta +it)|^2$, a quantity whose large-$t$ behavior has been extensively studied as a diagnostic of quantum chaos, particularly in the holographic context \cite{Polchinski}. The landmarks of quantum chaotic behavior in
$\zeta(\beta, t)$ are the `ramp' and the `plateau', which are sensitive to correlations in the discrete structure of energy eigenvalues. Since black-hole gravitational backgrounds miss the discreteness of the spectrum (cf. \cite{brab}), we do not expect the black hole saddle to contribute significantly to the ramp and plateau regimes.\footnote{We should not confuse the notion of `black-hole saddle' discussed here with the related, but different, double-cone wormhole saddles \cite{SSS} which are conjectured to characterize the ramp of the SFF.} However, it is expected that it could contribute to the `slope' describing the decrease of the spectral form factor for $0<t < t_{\rm dip}$, where $t_{\rm dip}$ marks the beginning of the ramp (typically a long time, scaling with a power of $N_*$).   

In this note, we  notice the following peculiar fact: the Euclidean action of the smooth `cigar' manifold $X_\beta$ with spherical horizon and high temperature  behaves as follows under  analytic continuation  to complex temperatures $\beta \rightarrow \beta +i t$ 
\begin{equation}\label{pec}
I(X^\pm_\beta )\rightarrow N_* \, C^\pm_d \,\frac{t}{\ell} \;, \;\;{\rm for}\;\; t \gg  \ell \gg \beta\;,
\end{equation}
with $\pm$ denoting the large (small) black holes present at  temperatures $T=1/\beta\gg 1/\ell$, that is, high compared to the  inverse radius of curvature $\ell$ of AdS$_{d+1}$.  The constants $C^\pm_d$ are imaginary for even $d$ and real for odd $d$. In this case, $C^-_d$ is negative for $d \equiv 3\; ({\rm mod} \;4)$ and
$C^+_d$ is negative for $d\equiv5 \;({\rm mod}\; 4)$. This would imply that the semiclassical prediction for the slope of the spectral form factor of a CFT$_d$ with gravity dual  \cite{Polchinski} would {\it diverge exponentially} on any sphere of dimension $4$ $({\rm mod} \;4)$!
\begin{equation}\label{bad}
|Z(\beta +it)|^2  \approx e^{-2 \,{\rm Re}\;I^+_{\beta +it}} \sim e^{+ 2\, N_* |C^+_d | |t /\ell |} \;, \;\; {\rm for}\;\;\; |t/\ell | \rightarrow \infty\;.
\end{equation} 
Even if we stay at $d\equiv3\,({\rm mod}\,4)$, the very negative action of the small black-hole saddle may induce us to think that it could dominate the path integral at large $t$. Such behavior is incompatible with physical expectations. 

A related problem is that the semiclassical approximation to the inverse Laplace transform, computing the microcanonical density of states 
\begin{equation}\label{ilp}
 \Omega(E)=\frac{1}{2 \pi i}  \int_{\Gamma:=\beta_0+i \mathbb{R}} d\beta\, Z(\beta)\, e^{\beta E},
\end{equation}
would be ill-defined in $d\equiv5$ (mod $4$), even for the `safe' saddle points representing large AdS black holes, if we simply plug in the analytic continuation of the black hole saddle. This is happening despite the fact that the local quadratic analysis around saddle points gives sensible results \cite{Barbon_2025}, even coping with negative modes around unstable saddles \cite{gpy}. 

On the other hand, we know that such pathologies must be artifacts of the careless  analytic continuation,  because the complex-temperature partition function admits a uniform bound 
provided the real-temperature partition function is well defined:
\begin{equation}\label{bound}
 \Big\vert Z(\beta+i t)\Big\vert = \Bigg\vert  \int dE\,e^{-(\beta+i t) E}\,\Omega(E)\Bigg\vert \leq  \int dE\,\Big\vert e^{-(\beta+i t) E}\Big\vert\,\Omega(E) = Z(\beta)\;.
\end{equation}
\\
In this work, we discuss a prescription that is able to reconcile the success of the local study of saddle points with the tension provided by the divergences of $Z(\beta+i\,t)$. This prescription amounts to the usual extension of $X^\pm_\beta$ to manifolds with a conical singularity, corresponding to thermal manifolds of `off shell' black holes, in which we take the mass parameter $M$  and the Euclidean inverse temperature $\beta$  as independent variables. This leads to a form of the Euclidean action in two independent real variables $I(M, \beta)$, and one  then considers the integral
\begin{equation}\label{bint} 
Z(\beta) = \int  d\mu(M) \,e^{-I( \beta,M)} \;,
\end{equation}
where $d\mu(M)$ is a physically motivated measure over real values of $M$ at fixed $\beta$. Using the analytic continuation of $I( \beta+it, M)$ into the complex $M$ plane, we analyze the contour deformations
that capture the leading contributions at different values of $t$. We find that this two-step procedure removes the pathologies found previously, since the saddles with very negative action do not lie in the relevant contour of integration for  $t \gg \beta$. 

This paper is organized as follows. In Section 2, we describe the phenomenon of analytic blow-up at large $t$. In Section 3, we define the model integral over off-shell black-hole manifolds and consider its evaluation by saddle point methods. Section 4 lists our concluding comments and outlook.

\section{Analytic Divergences of Black-Hole Partition Functions}

In this section, we study the behavior of the Euclidean action $I(\beta)$ of AdS black holes as a function of inverse temperature $\beta$, at large values of ${\rm Im}(\beta)$. The Euclidean action is defined by the usual expression involving the Einstein--Hilbert action corrected by the Gibbons--Hawking--York term, 
\begin{equation}\label{eac}
I(X) = -{1\over 16\pi G} \int_X (R-2\Lambda) - {1\over 8\pi G} \int_{\partial X} {\rm tr} \, K + {\rm counterterms} \;,
\end{equation}
and appropriate counterterms to render it finite from large-volume divergences. In $d+1$ bulk dimensions we use the conventions $\Lambda = -d(d-1)/2\ell^2$, with $\ell$, the radius of curvature of AdS$_{d+1}$, as well as the radius of the $(d-1)$-dimensional sphere where the CFT$_d$ is defined. The metric takes the form
\begin{equation}\label{metric}
ds^2 = f(r)\,d\tau^2 + {dr^2 \over f(r)} + r^2 d\Omega^2
\end{equation} 
with $\tau \equiv \tau + \beta$ and 
\begin{equation}\label{proff} 
f(r) = 1 + {r^2 \over \ell^2} - {16\pi G M \over (d-1) \Omega_{d-1} \, r^{d-2}} \;.
\end{equation} 
Here, $M$ is the mass of the black hole and $\Omega_{d-1}$ is the volume of the unit ${\bf S}^{d-1}$ sphere. 
The mass $M$ can be traded by the horizon radius $r_s$, the largest root of $f(r_s)=0$, and is tied to the inverse temperature by demanding smoothness at the horizon, yielding the relation 
\begin{equation}\label{ht}
    \beta ={4\pi \,r_s\,\ell^2 \over d\, r_s^2  + (d-2)\ell^2} \;,
\end{equation}
which we may solve as \footnote{Here, and in what follows, we take the branch cuts of power functions and logarithms running along the negative real axis.}
\begin{equation}\label{rofbeta}
    r_s = {{\bar r} \over \beta} \left( {\bar \beta} \pm \sqrt{{\bar \beta}^2 - \beta^2} \right)\;,\qquad {\bar r} \equiv \ell \sqrt{d-2 \over d}\;, \qquad {\bar \beta} \equiv {2\pi \over d-2} \,{\bar r}\;.
\end{equation}

With these preliminaries, the final answer for the Euclidean action can be written as 
\begin{equation}
    I(\beta) = \beta\,M(\beta) - S(\beta)\;,
\end{equation}
where $M(\beta)$ is the ADM mass of the black hole, with $r_s(\beta)$ solved via (\ref{ht}) 
\begin{equation}
    M= {(d-1) \Omega_{d-1} \over 16\pi G} \,r_s^{d-2}\,\left(1 + \frac{r_s^2}{\ell^2}  \right) \;,
\end{equation}
and $S(\beta)$ is the entropy 
\begin{equation}
    S(\beta) = {\Omega_{d-1} \over 4G} \,r_s^{d-1} \;.
\end{equation}
In what follows, we work in a unit system in which  $\ell=1$. 

We are now ready to study the analytic continuation of these expressions. 
We  assume $\beta \ll 1$ so that both solutions exist for real parameters. As we give $\beta$ a positive imaginary part $\beta \rightarrow \beta +it$, $t>0$, the
horizon radius $r_s$ gains a negative imaginary part for the case of the large AdS black hole ($+$) and a positive imaginary part for the case of the small AdS black hole ($-$). Taking $t\gg \ell$ we find
\begin{equation}\label{lim}
r_s^{(\pm)} \rightarrow \mp i\,{\bar r}\;, \qquad t \rightarrow +\infty\;.
\end{equation} 
Therefore, the entropy term in the Euclidean action goes to a constant as $t\rightarrow \infty$, as well as the ADM mass. Since the contribution of the ADM mass
is multiplied by $\beta+it$, this is the dominant term at large $t$, resulting in
\begin{equation}\label{asym}
I(\beta +it) \rightarrow (\beta +it) {(d-1)\Omega_{d-1} \over 16\pi G} \left(1- {\bar r}^2  \right) \,(\mp i \,{\bar r})^{d-2}\;,
\end{equation}
so that, keeping track of the signs, we find the result quoted in the introduction 
\begin{equation}\label{res}
I(X^\pm) \rightarrow N_*\,C^{\pm}_d \,t 
\;,
\end{equation}
where 
\begin{equation}\label{cs}
C^\pm_d = \mp\, \,{2 \over d}\,{\bar r}^{\,d-2} \,(\mp i)^{d-1} \;,
\end{equation}
and  we have defined 
\begin{equation}\label{nstar}
N_* \equiv {\Omega_{d-1} (d-1) \over 16\pi G} \;,
\end{equation}
as the parameter controlling the effective number of field species in the CFT, proportional to the central charge. 

The conclusion is that the saddle value of the partition function diverges exponentially at large $t$ for $d\equiv 5 \,({\rm mod} \,4)$ in the case of the large AdS black holes and is exponentially suppressed for other odd values of $d$. The opposite behavior occurs for small black holes, namely the saddle value of the partition function now diverges for $d\equiv 3 \,({\rm mod}\,4)$ and is exponentially suppressed for the rest of odd values of $d$. When $d$ is even, both cases give bounded oscillatory saddle contributions to the partition function. 

In assessing the significance of these results, we should perhaps distinguish between the direct interpretation of $Z(\beta + it)$ as the holomorphic square-root of the SFF, and its  use within the inverse Laplace transform (\ref{ilp}) determining the density of states. While constant and oscillating asymptotics may be considered
tolerable in (\ref{ilp}), given that $\Omega(E)$ is in general a distribution, anything different from a decay in time is certainly not a standard behavior for the `slope' of the SFF. Of course, it is to be expected that, as $\beta +it$ explores the complex plane, the black-hole saddles may lose dominance of the path integral but, for those values of $t$ for which a saddle gives an exponential enhancement of  $\exp (-I(\beta +it))$, that particular saddle cannot remain within 
 the integration cycle defining $Z(\beta +it)$. Our analysis in the following section drives this point home in a simplified model of the partition function.

\section{The Model Integral} 

In order to study whether a given saddle contributes to the integration cycle of the gravitational path integral, we need some concrete ansatz for this path integral. A maximally conservative attitude is to declare that the gravitational path integral is intrinsically semiclassical and only makes sense as a perturbative expansion around a saddle point. While this prescription may be pragmatically correct, it is true that some knowledge of the integration measure away from the saddle is needed in order to specify the perturbation technique.   We will adopt here the standard `minisuperspace' prescription, which can be considered a minimal extension of the above, in which one specifies a particular finite-dimensional integration contour that is physically motivated, and regards the perturbative expansion as integrating over the orthogonal directions in field space. In particular, we consider a family of black-hole manifolds corresponding to `off shell' black holes, in which we relax the constraint (\ref{ht}) between the temperature and the horizon radius, thereby introducing a conical defect at the horizon. The action $I(\beta, r_s)$ becomes now a function of two parameters given by the same expressions as before, 
\begin{equation}\label{offshell}
I(\beta, r_s) = \beta \,M(r_s) - S(r_s)\;,
\end{equation}
with the ADM mass and BH entropy expressed as functions of $r_s$. This can be argued by cutting the complete manifold $X(\beta, r_s)$ into two pieces: a small $\epsilon$-sized piece containing the conical singularity, with topology of a disk times the horizon sphere, and the remainder with cylindrical topology. The first term evaluates to $-S(r_s)$ despite the conical defect, by the Gauss--Bonnet theorem, and the second term computes $\beta M_{\rm ADM}$ as explained in \cite{Banados}.\footnote{The allowance of conical singularities  can be argued as a natural prescription from a Lorentzian formulation of gravitational path integrals that is able to deal with the conformal factor problem \cite{Marolf, Kolanowski:2026gii}.}

In the limit $r_s \rightarrow 0$, the singular manifolds $X(\beta, r_s)$ are assigned zero action, and the black hole size is formally vanishing. Therefore, this limit recovers the vacuum AdS manifold with thermal topology. From this point of view, we can see the off-shell black hole manifolds as interpolating between two saddles: the vacuum manifold and the true black hole manifold $X(\beta)$.  This interpolation is not smooth, because there is a topological flop at $r_s=0$, which is expected to be regularized by stringy effects in a Euclidean version of the correspondence between black holes and long strings \cite{HorPol} (see also \cite{BRH}).  Pragmatically, we can define the family of manifolds $X(\beta, r_s)$ as given by the off-shell black holes of mass $M(r_s)$ for $r_s >\sqrt{\alpha'}$ and as the vacuum AdS manifold decorated with excitations of energy $M(r_s)$ for $r_s < \sqrt{\alpha'}$. In a low-energy effective field theory description, black holes with $r_s < \sqrt{\alpha'}$ are integrated out and replaced by higher-derivative effective operators on the dynamics of the vacuum manifold.  

Defining a dimensionless variable $x=r_s /\ell$ we are led to the integral
\begin{equation}\label{minisup}
    Z(\beta)\approx  \int_{x_0}^\infty d\mu(x)\, e^{-I(\beta, x)} \;,
\end{equation}
where $I(\beta, x)$ is defined as in (\ref{offshell}), corrected by a tower of higher-dimension operators and loop effects of light degrees of freedom. This will involve the usual thermal gases of light states, such as gravitons, but also possibly thermal gases of long strings and logarithmic corrections due to approximate zero modes.\footnote{The simplest example being the term $-\log (\ell^2 M / 2\pi \beta)^{d/2}$ corresponding to the free energy in the center of mass degree of freedom of a small black hole, approximated as a particle of mass $M$ in a $d$-dimensional box of size $\ell$. Other examples include the Schwarzian modes in near-extremal throats.}
The measure $d\mu(x)$  is determined by interpreting (\ref{minisup}) as a proxy for the CFT integral 
\begin{equation}\label{genc}
Z(\beta) = {\rm Tr}\, e^{-\beta H} = \int_{0}^\infty dE \,\Omega (E) \,e^{-\beta E},
\end{equation} 
where $\log \Omega(E)$ is taken to be proportional to the Bekenstein--Hawking entropy. If we define the microscopic entropy at energy $E$ as the logarithm  of the number of states below $E$, then the measure is $d\mu = dx \,dS/dx$ for any integration parameter $x$. The infrared cutoff $x_0$ is related to the energy scale at which the sum over CFT states ceases to be well-approximated by a continuous integral. If we enrich the dynamics with graviton degrees of freedom and other propagating excitations (arising at one loop and beyond), we may take $E_0 \sim 1/\ell$ as the gap of the CFT on the sphere of radius $\ell$. More conservatively, we may set $x_0$ at the scale where black holes start dominating the density of states.  In a first approximation in which we keep only the leading terms in the large-$N_*$ expansion, we can neglect both the measure Jacobian and the perturbative loop contributions. In the same vein, given that the energy bands dominated by black hole states have densities of states
with exponential dependence on $N_*$, the energy thresholds where the density of states loses black-hole dominance are suppressed by inverse powers of $N_*$ in the $x$ variable. Therefore, we can extend the $x$ contour of integration to the full naive domain and leave for later analysis the `boundary layer' dynamics at the low-energy endpoint. 

Incidentally, the form (\ref{minisup}) is also well-suited for more radical redefinitions of the partition function, such as, for example, a restriction to a narrow microcanonical band around the saddle point of interest. This can be done by adjusting the measure $d\mu$ accordingly. Inserting a gaussian of sufficiently narrow energy width $\Delta$ around, say the large black-hole saddle $x_B$, corrects the saddle value $\exp(-I_B)$ by a factor of $\exp(\beta^2 \Delta^2 /2)$. Upon analytic continuation $\beta \rightarrow \beta +it$, this gives a Gaussian damping term for $t\gg \Delta^{-1}$ which regularizes any possible  pathological behavior in $I_B$. Our purpose in the rest of the paper is to argue that the large $t$ behavior of $Z(\beta +it)$ is still well defined without the need for an explicit microcanonical window. 

\subsection{ Picard--Lefschetz Analysis of the Black-Hole Band}
\begin{figure}[htbp]
   \centering
     \includegraphics[width=4in]{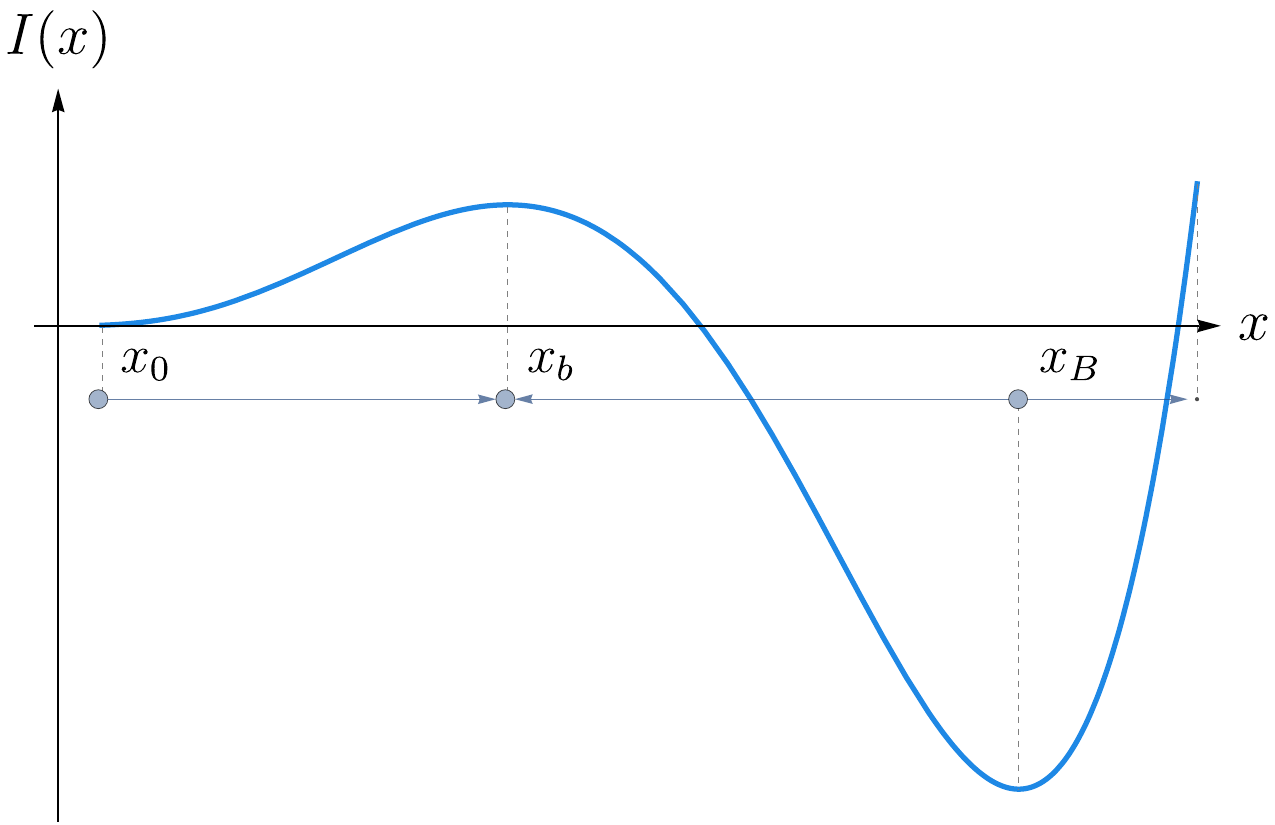}
\caption{Plot of the Euclidean action as a function of the horizon radius $x$ for a fixed real value of $\beta\ll1$. The maximum of the action is given by the small black hole saddle $x_b$, while the minimum corresponds to the large black hole saddle $x_B$. We use $x_0$ to represent the endpoint of the integration cycle $\mathcal{C}_0$. }
\label{Fig:1}
\end{figure}

\begin{figure}[htbp]
   \centering
   \includegraphics[width=4.65in]{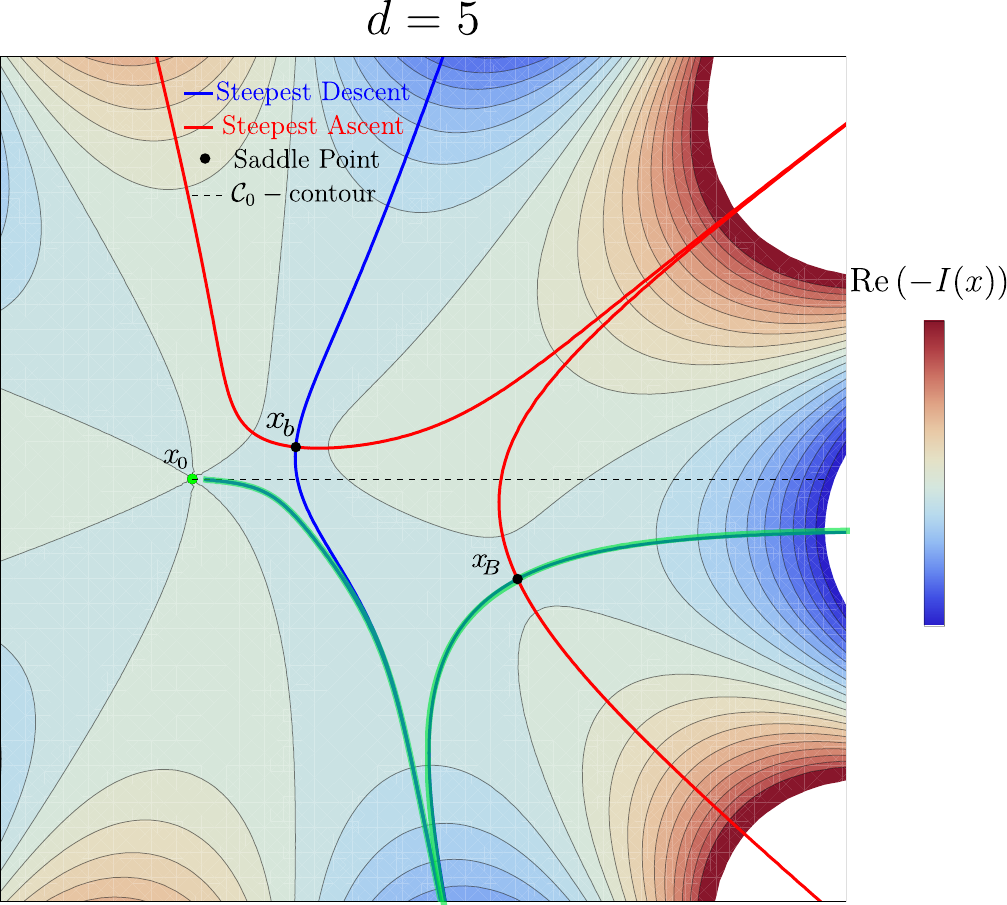}
\caption{ 
Contour plot for the real part of $-I(\beta+it\,,x)$. The two saddle points $x_b,\,x_B$ migrate into the complex plane, going to the imaginary axis as $t$ increases. Steepest descent paths are depicted in blue, while steepest ascents are represented in red. For values of $t<t^{(B)}_S$, of order $\beta$ in $d>3$, and order $\ell$ in the special $d=3$ case, the contour defining the integral for the partition $\mathcal{C}_0$, can be decomposed into an endpoint piece $\mathcal{J}_0$ and the large black hole saddle $\mathcal{J}_B$. The contributing parts are highlighted in green. The thimble corresponding to the small black hole $\mathcal{J}_b$ does not enter into the decomposition of the $\mathcal{C}_0$ contour.
}
\label{Fig:2}
\end{figure}

\noindent
We now consider the integral (\ref{minisup})  and assume high temperatures
$\beta \ll \ell= 1$ to ensure the dominance of the large AdS saddle of the partition function at real temperatures.\footnote{See \cite{Mahajan} for a discussion of complex saddles at low temperatures. This analysis was further developed in \cite{Ailiga}, where the connection with the Lorentzian formalism was made explicit. }   In studying the dynamics of the black hole saddles with action of $O(N_*)$, we can neglect in a first approximation all other physical thresholds in the spectrum of the strongly coupled CFT, whose action will be suppressed  by inverse powers of $N_*$, compared to that of the black holes. In particular, this means  that we can set $x_0 \approx 0$ to this order of accuracy. Furthermore, the measure factor gives contributions of $O(\log N_*)$ in the exponent so that, at large $N_*$,  we may consider the simplified problem 
\begin{equation}\label{int}
Z(\beta +it) \approx \int_0^\infty dx \,e^{-I( \beta+it,\,x)} \;,
\end{equation} 
where 
\begin{equation}\label{ac}
I( \beta+it,\,x) = (\beta +it) \,N_* \,x^{d-2}\, (1+ x^2) - {4\pi \over d-1} N_* \, x^{d-1} \;.    
\end{equation}
The defining contour  $\mathcal{C}_0 = [0, \infty)$ runs through the positive real axis. For $t=0$ we have two saddle points given by 
\begin{equation}\label{saddles}
x_s = {2\pi \pm \sqrt{ 4\pi^2 -d(d-2)\,\beta^2\,} \over d\, \beta} \;,
\end{equation}
with the $\pm$ signs corresponding to the large ($x_B$) and small ($x_b$) black holes respectively. In the high-temperature region, $\beta \ll 1$, the large black-hole saddle is a maximum of $-I$  `far out' the real axis, whereas the small black-hole saddle is a minimum of $-I$ sitting `near' the origin (cf. \cref{Fig:1}): 
\begin{equation}\label{farnear}
x_B \approx {4\pi \over d \,\beta} - {d-2 \over 4\,\pi} \beta \;, \qquad x_b \approx {d-2 \over 4\pi} \beta\;.
\end{equation}
The breakup of the defining contour $\mathcal{C}_0$ into thimbles needs regularization, since the saddles $x_B$ and $x_b$ sit on a Stokes line for real $\beta$.\footnote{For an introduction to Picard-Lefschetz theory applied to physics see e.g. \cite{WittenAMS}.}
The deformation of interest $\beta \rightarrow \beta +it$ can be used to define the thimble decomposition, starting with a very small $t$. In this case, the
large black-hole saddle $x_B$ picks a small negative imaginary part, whereas the small black hole saddle $x_b$ moves slightly upwards from the real axis into the positive imaginary domain, see \cref{Fig:2}. One finds that the thimble of $x_b$ cuts the original contour and attracts the thimbles of both the large black hole saddle $x_B$ and the endpoint (see \cite{malturiaci} for similar considerations in a different context). 

Let us denote $\phi = {\rm arg}\, (\beta +it)$. Then, the asymptotic valleys of $ -I$ are determined by maximizing ${\rm Re} \,I$ as $|x| \rightarrow \infty$. This results in the directions ${\rm arg}\,(x) \equiv -\phi / d \;\; ({\rm mod}\;2\pi /d)$. Hence, the thimbles of both the endpoint and the $x_B$ saddle are asymptotic to ${\rm arg}\,(x) = -\phi/d - 2\pi /d$ and the contour $\mathcal{C}_0$ breaks up into the homologically equivalent  steepest descent contour $\mathcal{J}_0 + \mathcal{J}_B$ where $\mathcal{J}_0$ starts at the origin and descends towards the valley ${\rm arg}\,(x) = - (\phi + 2\pi)/d$ and $\mathcal{J}_B$ emerges from that valley and passes through $x_B$ on its way to the asymptotic line $\arg\,(x) = -\phi/d$. This is shown in \cref{Fig:2}, where activated thimbles are highlighted in green. On the other hand, the thimble of the small black hole, $\mathcal{J}_b$, does not participate in the  homology class of $\mathcal{C}_0$. One way to see this is to consider the ascent thimble passing through $x_b$, which starts in the direction ${\rm arg}\,(x-x_b) \equiv  -(d-3)\phi/2 \;({\rm mod} \,\pi)$, and follow the asymptotic lines that minimize ${\rm Re}\,I$ as $|x| \rightarrow \infty$, given by ${\rm arg}\,(x) \equiv (\pi-\phi)/d \; ({\rm mod}\; 2\pi /d)$. Since this ascent thimble does not intersect $\mathcal{C}_0$, the $x_b$ saddle remains topologically disconnected. 

Next, we proceed to explore the complexified temperature up to $t = O(\beta)$. In this regime, it is
still true that $|\beta +it | \ll 1$ and the asymptotic forms of (\ref{farnear}) are still valid with the substitution $\beta \rightarrow \beta +it$. A useful exact expression to analyze the value of the action at the saddles is
\begin{equation}\label{sval}
 I_s = -{N_* \over d-1} \,(\beta+i\,t) \,\left(x_s^d -x_s^{d-2}\,\right)\;,
\end{equation}
from which we can deduce the asymptotic forms of the saddle values
\begin{equation}\label{asys}
    I_B \approx -\frac{N_*}{d-1} \left({4\pi \over d}\right)^d \,(\beta+it)^{1-d}\;, \qquad I_b \approx \frac{N_*}{d-1}  \left({d-2 \over 4\pi}\right)^{d-2} \, (\beta +it)^{d-1}\;. 
\end{equation}
These estimates can be used to determine the dominant contribution along the integration contour, as well as the pattern of Stokes jumps, which eventually changes the topology of the contours. 

The pattern of dominance is controlled by the real part of the action, whose contributions from the saddles and the endpoint have a `triple point' at $\phi_D = \pi/2(d-1)$ at which $0={\rm Re}\,I_0 = {\rm Re}\,I_B = {\rm Re}\,I_b$. For $t > \beta \;\tan(\pi/2(d-1))$, the large black-hole saddle $x_B$ stops giving the dominant contribution, and the integral becomes dominated by the endpoint at the origin. Notice that the dominant critical point becomes $x_b$, but at this stage, the small black hole is still  disconnected from the linear combination of contributing thimbles. 

\begin{figure}[htbp!]
   \centering
   \includegraphics[width=4.65in]{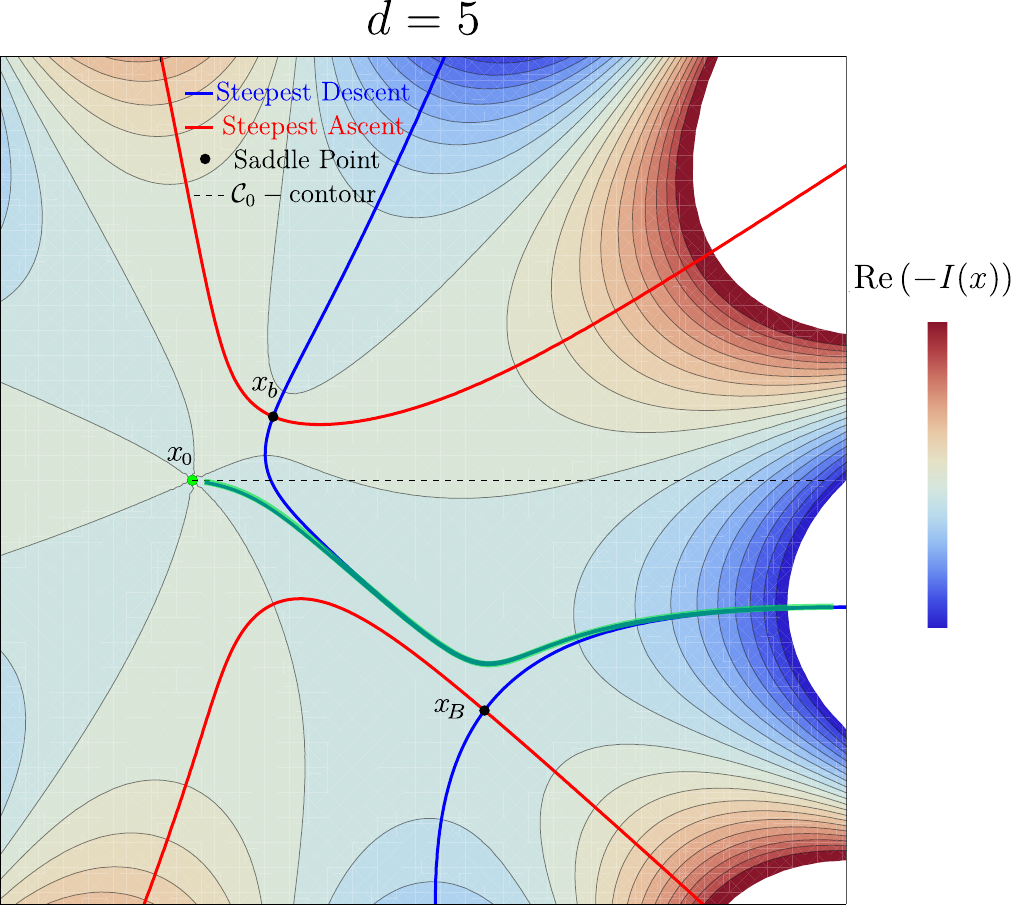}
\caption{
Contour plot for the real part of $-I(\beta+it\,,x)$. The two saddle points $x_b,\,x_B$ migrate into the complex plane, going to the imaginary axis as $t$ increases. Steepest descent paths are depicted in blue, while steepest ascents are represented in red. For values of $t>t^{(B)}_S$, of the order of $\beta$ in $d>3$ and order $\ell$ in the special $d=3$ case, a Stokes transition takes place. 
Across the Stokes line, the endpoint steepest descent contour jumps according to $\mathcal{J}_0\to\mathcal{J}'_0=\mathcal{J}_0+\mathcal{J}_B$. After the transition, the contour defining the integral for the partition function, highlighted in green, satisfies $\mathcal{C}_0\equiv \mathcal{J}'_0$, and the large black hole saddle is no longer picked up by the integration contour. The thimble corresponding to the small black hole $\mathcal{J}_b$ does not enter into the decomposition of the $\mathcal{C}_0$ contour.
}
\label{Fig:3}
\end{figure}

On the other hand, the topological pattern of contours does change when Stokes lines occur between critical points. This is controlled by the imaginary part of the action, and again, within the approximations (\ref{asys}),  we find a `triple point' at
$\phi_S = \pi/(d-1)$ at which $0 = {\rm Im}\,I_0 = {\rm Im}\,I_B = {\rm Im}\,I_b$.
Beyond this point, the contour abandons the valley of the small black-hole thimble and flows directly from the origin to the asymptotic valley in the direction ${\rm arg}\,(x) = - \phi/d$, as shown in \cref{Fig:3}. This contour, which we denote $\mathcal{J}'_0$, is in the same homology class as the original contour $\mathcal{C}_0$. 

Therefore, we conclude that, for $t>t_S= \beta \;\tan\,(\pi /(d-1))$, the large AdS black hole saddle disconnects topologically from the relevant sum of thimbles. To be more precise, this happens as long as $d>3$. 

\begin{figure}[htbp]
    \centering
        \centering
        \includegraphics[width=0.8\textwidth]{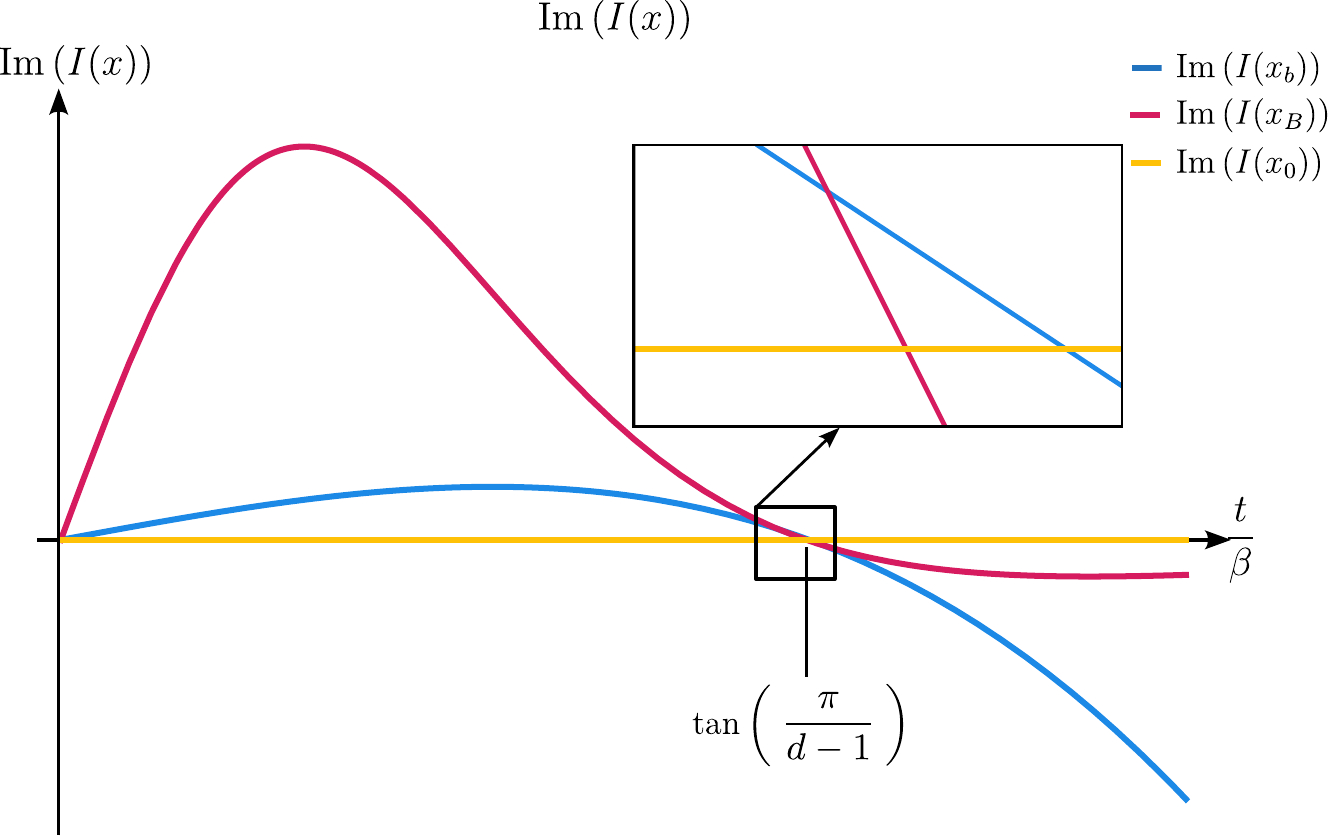}
    \caption{Imaginary part of the action $I(\beta+it,\,x)$, as a function of $t/\beta$ evaluated at the critical points $x_b$, $x_B$ and at the endpoint $x_0$ for $d=5$ and $\beta=0.1$. For visualization purposes, the action at $x_b$ had to be rescaled.  The value $t/\beta=\tan(\pi/(d-1))$, corresponding to the degenerate triple point in the $\beta\to0$ limit, is also represented. In the box, we present a zoom of how the triple point is resolved by $O(\beta^2/\ell^2)$ corrections.}\label{Fig:4.1}
\end{figure}

\begin{figure}[htbp]
    \centering
        \centering
        \includegraphics[width=0.8\textwidth]{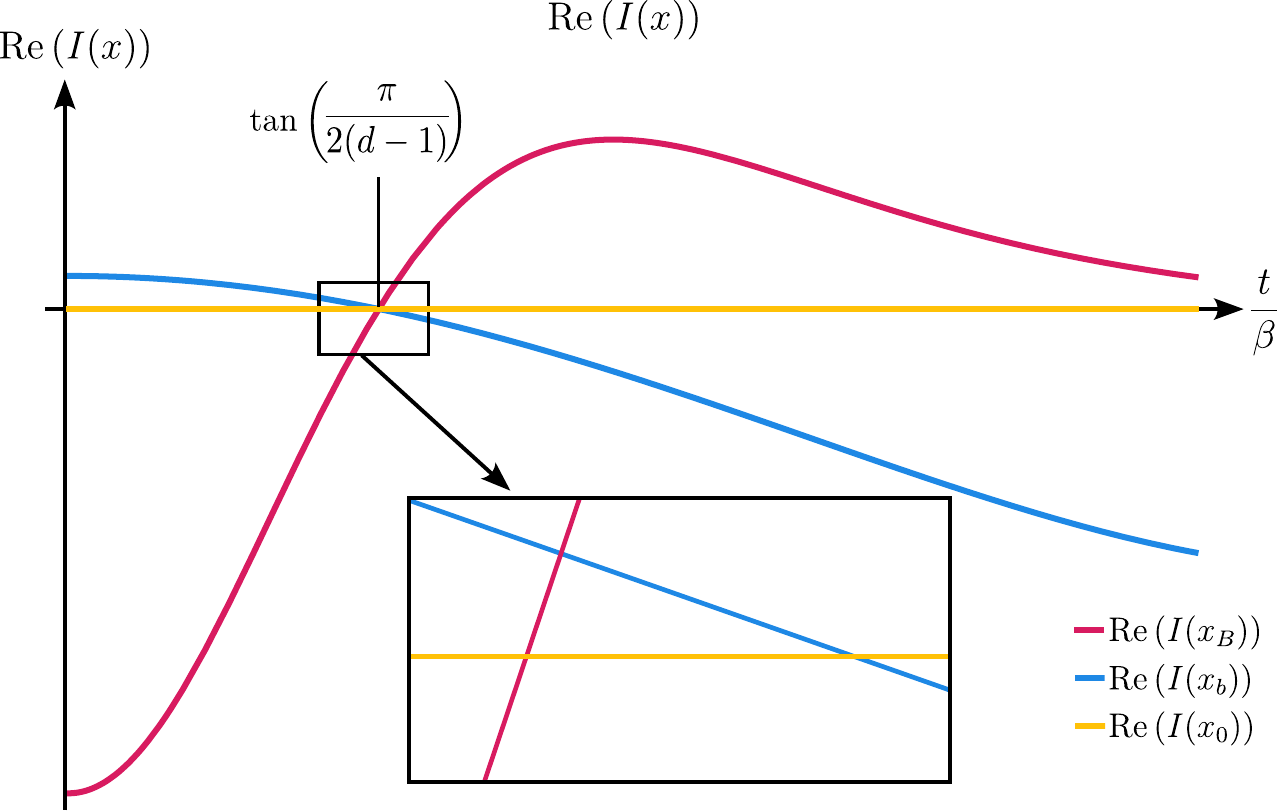}
    \caption{Real part of the action $I(\beta+it,\,x)$, as a function of $t/\beta$ evaluated at the critical points $x_b$, $x_B$ and at the endpoint $x_0$ for $d=5$ and $\beta=0.1$. For visualization purposes, the action at $x_b$ had to be rescaled. The value $t/\beta=\tan(\pi/2(d-1))$, corresponding to the degenerate triple point in the $\beta\to0$ limit, is also represented. In the box, we present a zoom of how the triple point is resolved by $O(\beta^2/\ell^2)$ corrections.}\label{Fig:4.2}
\end{figure}

\begin{figure}[htbp]
\centering

\begin{tikzpicture}[scale=1.15]

\begin{scope}[shift={(0,0)}]
\node at (-1,3.4) {\large (i)};

\fill (1.6,3.2) circle (2.5pt);
\node[above] at (1.6,3.2) {$x_B$};

\draw[flow, line width=2pt] (0.2,0.2) to[out=80,in=190] (1.63,3.23);
\draw[flow, line width=2pt] (1.6,3.23) to[out=-10,in=100] (2.95,0.2);

\fill (0.8,1.6) circle (2.5pt);
\node[above right] at (0.8,1.6) {$x_b$};

\draw (0.8,1.6) to[out=190,in=30] (0.47,1.5);
\draw (0.33,1.42) to[out=230,in=85] (-0,0.2);
\draw (0.8,1.6) to[out=-10,in=165] (2.75,1.27);
\draw (2.9,1.24) to[out=-10,in=100] (4.2,0.2);

\fill (-0.8,2.4) circle (2.5pt);
\node[above right] at (-0.8,2.4) {$x_0$};

\draw[flow, line width=2pt] (-0.8,2.4) to[out=-10,in=100] (-0.1,0.2);

\draw[line width=1.75pt] (-0.25,0.17)--(-0.25,0)--(0.3,0)--(0.3,0.17);
\draw[line width=1.75pt] (2.8,0.17)--(2.8,0)--(3.1,0)--(3.1,0.17);
\draw[line width=1.75pt] (4.1,0.17)--(4.1,0)--(4.3,0)--(4.3,0.17);

\node[below] at (1.7,-0.1) {$\mathcal{C}_0\equiv{\mathcal{J}^{t<t_D}_0}+\boxed{\mathcal{J}^{t<t_D}_B}$};
\end{scope}

\begin{scope}[shift={(5.5,0)}]
\node at (-1,3.4) {\large (ii)};

\fill (1.6,1.6) circle (2.5pt);
\node[above] at (1.6,1.6) {$x_B$};

\draw[flow, line width=2pt] (0.17,0.2) to[out=80,in=190] (1.63,1.63);
\draw[flow, line width=2pt] (1.6,1.63) to[out=-10,in=100] (2.95,0.2);

\fill (1,3.2) circle (2.5pt);
\node[above right] at (0.8,3.2) {$x_b$};

\draw (1,3.2) to[out=195,in=85] (-0,0.2);
\draw (1,3.2) to[out=-5,in=100] (4.2,0.2);

\fill (-0.8,2.4) circle (2.5pt);
\node[above right] at (-0.8,2.4) {$x_0$};

\draw[flow, line width=2pt] (-0.8,2.4) to[out=-10,in=100] (-0.1,0.2);

\draw[line width=1.75pt] (-0.25,0.17)--(-0.25,0)--(0.3,0)--(0.3,0.17);
\draw[line width=1.75pt] (2.8,0.17)--(2.8,0)--(3.1,0)--(3.1,0.17);
\draw[line width=1.75pt] (4.1,0.17)--(4.1,0)--(4.3,0)--(4.3,0.17);
\node[below] at (1.7,-0.1) {$\mathcal{C}_0\equiv\boxed{\mathcal{J}^{\,t_D<t<t^{(b)}_S}_0}+\mathcal{J}^{\,t_D<t<t^{(b)}_S}_B$};
\end{scope}

\begin{scope}[shift={(0,-4.8)}]
\node at (-1,3.4) {\large (iii)};

\fill (1.3,1.6) circle (2.5pt);
\node[above] at (1.3,1.6) {$x_B$};

\draw[flow, line width=2pt] (0.17,0.2) to[out=80,in=190] (1.33,1.63);
\draw[flow, line width=2pt] (1.3,1.63) to[out=-10,in=100] (2.95,0.2);

\fill (1,3.2) circle (2.5pt);
\node[above right] at (0.8,3.2) {$x_b$};

\draw (1,3.2) to[out=190,in=95] (3.13,0.2);
\draw (1,3.2) to[out=-5,in=100] (4.2,0.2);

\fill (-0.8,2.4) circle (2.5pt);
\node[above right] at (-0.8,2.4) {$x_0$};

\draw[flow, line width=2pt] (-0.8,2.4) to[out=-10,in=100] (-0.1,0.2);

\draw[line width=1.75pt] (-0.2,0.17)--(-0.2,0)--(0.3,0)--(0.3,0.17);
\draw[line width=1.75pt] (2.8,0.17)--(2.8,0)--(3.25,0)--(3.25,0.17);
\draw[line width=1.75pt] (4.1,0.17)--(4.1,0)--(4.3,0)--(4.3,0.17);
\node[below] at (1.7,-0.1) {$\mathcal{C}_0\equiv\boxed{\mathcal{J}_0^{\,t^{(b)}_S<t<t^{(B)}_S}}+\mathcal{J}^{\,t^{(b)}_S<t<t^{(B)}_S}_B$};

\end{scope}

\begin{scope}[shift={(5.75,-4.8)}]
\node at (-1,3.4) {\large (iv)};

\fill (1,1.6) circle (2.5pt);
\node[above left] at (1,1.6) {$x_B$};

\draw (0,0.2) to[out=80,in=190] (1,1.63);
\draw (1,1.63) to[out=-10,in=100] (2.75,0.2);

\fill (0.8,3.4) circle (2.5pt);
\node[above right] at (0.8,3.4) {$x_b$};

\draw (0.8,3.4) to[out=190,in=93] (3.1,0.2);
\draw (0.8,3.4) to[out=-5,in=100] (4.2,0.2);

\fill (-0.3,2.4) circle (2.5pt);
\node[above right] at (-0.3,2.4) {$x_0$};

\draw[flow, line width=2 pt] (-0.3,2.4) to[out=-10,in=105] (2.9,0.2);

\draw[line width=1.75pt] (-0.1,0.17)--(-0.1,0)--(0.2,0)--(0.2,0.17);
\draw[line width=1.75pt] (2.6,0.17)--(2.6,0)--(3.25,0)--(3.25,0.17);
\draw[line width=1.75pt] (4.1,0.17)--(4.1,0)--(4.3,0)--(4.3,0.17);

\node[below] at (1.7,-0) {$\mathcal{C}_0\equiv\boxed{\mathcal{J}^{t^{(B)}_S<t}_0}$};
\end{scope}

\end{tikzpicture}
\captionsetup{labelformat=empty}

\caption{Figures (i)--(iv): Sketch of the thimbles' evolution with $t$. Brackets denote convergence regions, and arrows represent the direction of integration for the contributing thimbles, represented with thicker lines. The decomposition of the integration contour in the relative homology is also included.
Height represents the value of $\operatorname{Re}\left(-I(x_i)\right)$, for $x_i\in\{x_0,\,x_b,\,x_B\}$. Boxes are used to indicate the dominance, in the saddle-point approximation, of the thimble corresponding to the highest saddle that is picked up by the integration contour. The diagrams show how the thimble for the small black hole $\mathcal{J}_b$ is never picked up by the contour. $t_D$ denotes the time at which, the $x_B$ thimble and the endpoint $x_0$ exchange dominance. $t^{(b)}_S$ and $t^{(B)}_S$ refer to the Stokes times of the correspondent saddle. }
\label{Fig:5}

\end{figure}
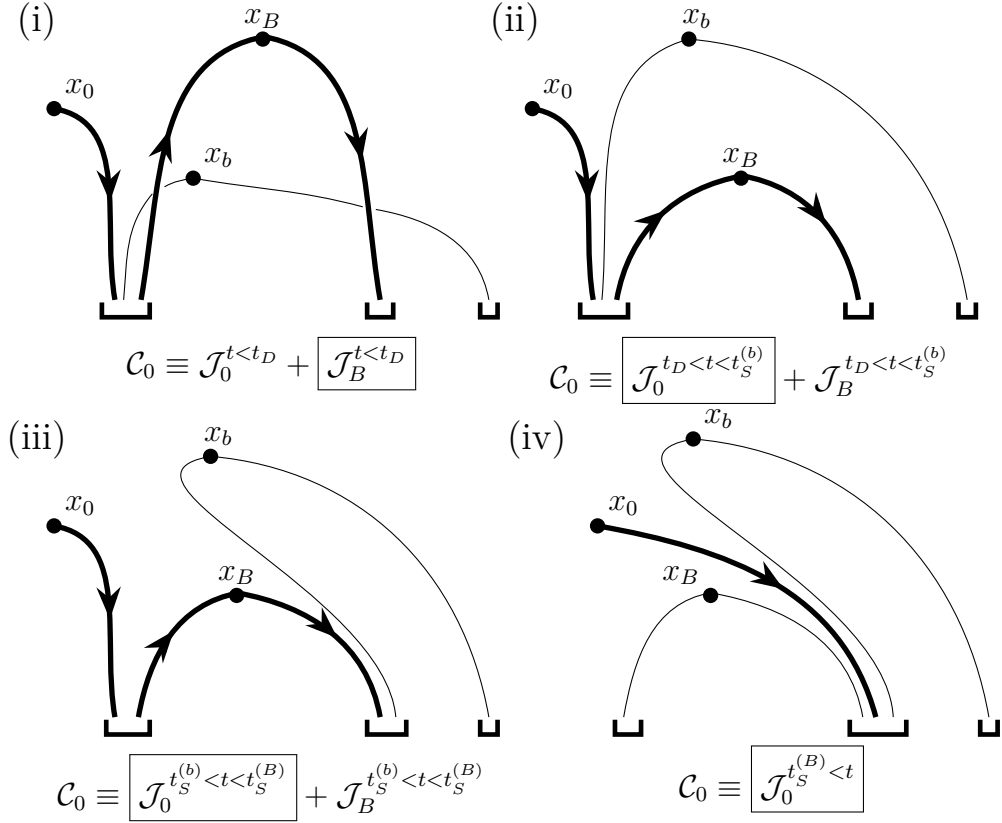

The precise pattern of contour rearrangements depends on how the subleading corrections break the triple point degeneracy of the leading approximation. To see this, we keep the next-to-leading terms in the high-temperature expansion of the saddles, i.e., we replace (\ref{farnear}) with 
\begin{align}
x_B &\approx {4\pi \over d (\beta+it)} \,\left(1-{d(d-2) \over (4\pi)^2} \,(\beta+it)^2 \right)\,,\label{nls1} \\
x_b &\approx {d-2 \over 4\pi} (\beta+it) \,\left(1+{d(d-2) \over (4\pi)^2} \,(\beta+it)^2 \right) \,.\label{nls2}
\end{align}
Evaluating now ${\rm Im} \,I_s$ to the next-to-leading order in the small $\varepsilon\equiv |\beta+it|$ expansion, one finds that the Stokes angles determined by ${\rm Im}\,I_s =0$ are shifted by
\begin{align}
    \phi_S^{(B)} &\approx {\pi \over d-1} - \varepsilon^2\,\left({d \over 4\pi}\right)^2  \,\sin \left({2\pi \over d-1}\right)\;, \label{shiftedS1}\\
    \phi_S^{(b)} &\approx {\pi \over d-1} -\varepsilon^2\,\left({d-2 \over 4\pi}\right)^2  \,\sin \left({2\pi \over d-1}\right)\;. \label{shiftedS2}
\end{align}
Notice that $\phi_S^{(B)} < \phi_S^{(b)}$, so that the detailed structure resolving the triple point is the one shown in \cref{Fig:4.1}: the first Stokes transition is between the two black hole saddles, and after that the large AdS saddle is the first to disconnect from the origin. The activated thimbles undergo the transition  $\mathcal{J}_0 + \mathcal{J}_B \rightarrow \mathcal{J}'_0$, with the small black-hole saddle $x_b$ remaining as a spectator, never connecting to the relevant set of contours. 
Actually, the Stokes line between the black hole saddles is determined by the condition $\operatorname{Im}I_b=\operatorname{Im}I_B$. Near the degenerate high-temperature Stokes angle, this differs from \cref{shiftedS1} only to higher order in $|\beta+it|$.\footnote{Indeed, $I_B=O(|\beta+i t|^{1-d})$ while $I_b=O(|\beta+i t|^{d-1})$, its displacement from the condition $\operatorname{Im}(I_i)=0$ is $O(|\beta+i t|^{2d})$.  }
\cref{Fig:4.2} shows a similar resolution of the dominance triple point at $\phi_D = \pi/2(d-1)$, which can be studied in an entirely similar perturbative analysis. The time evolution of the thimbles is represented in \hyperref[Fig:5]{Figures (i)--(iv)}


\subsubsection{Stokes phenomena in $d=3$.}
The pattern of Stokes phenomena in $d=3$ is special in the sense that, in the $\vert \beta+it\vert\ll1$ analysis, the Stokes line appears at $\phi_S\simeq\pi/2$, that is to say $t\gg\beta  $. The approximation indicates that the Stokes time cannot be $t_S=O(\beta)$.  In this case, however, one can still expand the action for $\beta\ll1$, for arbitrary $t-$values and determine the values $ t^{(b)}_S, \;t^{(B)}_S$ for which Stokes phenomena take place between the thimbles:
\begin{align}
\operatorname{Im}\left(I_0\left(\beta+i\,t^{(B)}_S\right)\right)&=\operatorname{Im}\left(I_B\left(\beta+i\,t^{(B)}_S\right)\right)\,,\\
\operatorname{Im}\left(I_B\left(\beta+i\,t^{(b)}_S\right)\right)&=\operatorname{Im}\left(I_b\left(\beta+i\,t^{(b)}_S\right)\right)\, .
\end{align}
Restoring the factors of $\ell$, we obtain 
\begin{align}
    t^{b}_S&\approx2\pi \ell\,\sqrt{\frac{2}{3}}\left(1-\frac{11}{48\pi^2}\frac{\beta^2}{\ell^2}+O\left(\frac{\beta^4}{\ell^4}\right)\right)\,,\\
    t^{B}_S&\approx2\pi \ell\,\left(1-\frac{5}{32\pi^2}\frac{\beta^2}{\ell^2}+O\left(\frac{\beta^4}{\ell^4}\right)\right)\,.
\end{align}
In the special case of $d=3$, we see that the dominance-transition time, which admits a perturbative analysis $t_D=O(\beta)\ll\ell$, is parametrically separated from the Stokes time ($t_S=O(\ell)$) corresponding to the topological recombination of the thimbles.

\subsection{Boundary Layer Analysis}
\noindent
The main message of the preceding  discussion is that,  as soon as $t$ becomes of order $\beta\ll 1$, the black-hole saddle loses dominance over the low-energy endpoint and, at slightly higher values of $t$, within the same order of magnitude, the black-hole saddle  disconnects altogether from the thimbles contributing to the semiclassical approximation (this requires $d>3$). At this stage, we revisit some of the simplifying assumptions that were laid down, such as the approximation $x_0 \approx 0$ and the lack of structure in the density of states at intermediate energy scales. 

A more detailed, yet still qualitative, picture of the density of states of strongly coupled CFTs with a gravity dual can be obtained by simply adding the universal massless excitations in the bulk, the gravitons, on top of the endpoint AdS vacuum manifold. Qualitatively, the entropy in an energy $E_g$ worth of gravitons scales like (still working in units $\ell=1$)  $S_g \approx s_g\, E_g^{d/(d+1)}$, with $s_g = O(1)$, related to the `radiation constant' in dimension $d+1$. One may consider adding an intermediate band of strings with Hagedorn spectrum, but a proper treatment of this system would require considering somewhat more realistic bulk manifolds featuring compact factors, an extension that we leave for future work. At any rate, we can obviate this phase by restricting the `high temperatures' determined by the choice of control parameter $\beta$  to still be sub-stringy, namely $\sqrt{\alpha'} \ll \beta \ll 1$. Furthermore, we must make sure that the microcanonical gas temperature never reaches the string scale. To implement this constraint, we consider the transition energy $E_{gb}$ between a graviton gas and a small black hole as determined by $S_{g}(E_{gb}) \sim S_b (E_{gb})$, or
\begin{equation}\label{match}
E_{gb}^{d \over d+1} \sim {1\over G} (G\,E_{gb})^{d-1 \over d-2}\;.
\end{equation}
Requiring the associated microcanonical temperature $\left(\partial S / \partial E\right)^{-1}$ to remain substringy at $E_{gb}$ gives the constraint 
\begin{equation}\label{nostring}
    {\sqrt{\alpha'}\over \ell} < {1\over N_*^{1\over 2d-1}}\;.
\end{equation}
An equivalent characterization of this bound is to say that the string scale cannot be arbitrarily larger than the Planck scale, namely
\begin{equation}\label{ns}
{\sqrt{\alpha'} \over \ell_{\rm Planck} } < N_*^{d \over (d-1)(2d-1)}\;.
\end{equation}

Another source of stringy corrections is the detailed `correspondence' between the small black holes and long strings \cite{HorPol} when the Schwarzschild radius becomes stringy, $r_s \sim \sqrt{\alpha'}$. This governs  the transition between the vacuum AdS manifold and the off-shell black hole manifolds and would show up in the low-energy effective action as a tower of higher-dimension operators. In this work, we will describe the bulk dynamics in the effective field theory approximation and drop the stringy corrections that would contribute to the fundamental integral (\ref{partf}) both at the level of the action and the measure.

We can write an ansatz for the entropy function in terms of the same variable $x$ that we use in the parametrization of the black hole states, by extending to the low-energy phases the analytic form of the mass of a small black hole, namely $E= N_* \,x^{d-2}$. This definition leads to
\begin{equation}\label{entropybands}
    S(x) = S_{\rm BH}  + S_{g} = {4\pi \over d-1} \,N_* \,x^{d-1}  + s_g \,N_*^{d/(d+1)} \,x^{d(d-2)/ (d+1)}\;. 
\end{equation}
For the parametric range of interest, $N_* \gg 1$, this ansatz presents the required qualitative behavior to interpolate between the two regimes of interest. The graviton gas band extends from the gap at $x_0 \sim 1/N_*^{1\over d-2}$ to the point where the small black-hole entropy begins to dominate, at $x_{gb} \sim 1/N_*^{1 \over 2d-1}$.

The resulting free energy generalizing the Euclidean action of the black holes is given  by the ansatz
\begin{equation}\label{blI}
I(x, \beta) = -{4\pi \over d-1} \,N_*\, x^{d-1}  - s_g \,N_*^{d/(d+1)} \,x^{d(d-2) \over d+1} + \beta \,N_* \,x^{d-2} \,(1+x^2)\;.
\end{equation}

In the regime of interest, $N_* \gg 1$, the critical points associated with the boundary-layer dynamics are geometrically well separated in the complex plane from the black-hole critical points. Therefore,  in discussing the graviton-gas band, it is convenient to drop the parametrization in terms of $x$, which was inspired by the horizon radius of a black hole, and use a normal energy variable $y = E \ell$. In terms of this, we can write a `boundary-layer' approximation to the partition function by keeping only the graviton  band with entropy $S_g (y) $: 
\begin{equation}\label{blz}
    Z(\beta +it)_{\rm BL} = \int_{y_1}^{\infty} dy \,{dS_g \over dy}\,e^{-I(y, \beta+it)_{\rm BL}}\;,
\end{equation}
where $y_1 = E_{\rm gap} \ell =1$ and the action is 
\begin{equation}\label{BLac}
    I(y, \beta+it)_{\rm BL} = (\beta +it)\, y -S_g (y) = (\beta + it)\, y  - s_g \,y^{d \over d+1}\;.
\end{equation}
We extend the integration to infinity in (\ref{blz}) despite the fact that the integrand, determined by $I_{\rm BL}$, is only a good approximation for $y\ll y_{gh} = N_* x_{gh}^{d-2} = N_*^{(d+1)/(2d-1)}$. As we will see, the dominant contributions to the integral are well within the `boundary layer' region $y\ll y_{gh}$.

We may also integrate by parts in (\ref{blz}) to write
\begin{equation}\label{intp}
    Z(\beta+it)_{\rm BL} = - e^{-(\beta +it) + s_g} + (\beta+it) \int_{1}^{\infty} dy\,e^{-I(y, \beta+it)_{\rm BL}} \;,
\end{equation}
and focus on the simpler integral, with $\gamma \equiv d/(d+1)$, 
\begin{equation}\label{zprima}
Z' = \int_1^\infty dy\,e^{\Phi(y)}\;, \qquad \Phi(y) = s_g \,y^{\gamma} - (\beta +it) \,y\;.
\end{equation} 

For $t=0$, the integral $Z'$ is dominated by the usual ideal  gas saddle, which moves into the lower complex plane as $t$ is turned on, 
\begin{equation}\label{gsad}
y_g = \left({\gamma \,s_g \over \beta + it}\right)^{d+1}\;.
\end{equation} 
The steepest descent contours from either the endpoint at $y=1$ or the gas saddle are asymptotic to the lines ${\rm arg}\,(y) \equiv -\phi \;({\rm mod}\,\,\pi)$. For small $t$, the original contour $\mathcal{C}_1 = [1, \infty)$ deforms 
into the thimble sum $\mathcal{J}_1 + \mathcal{J}_g$, where $\mathcal{J}_1$ is defined by the condition ${\rm Im}\,\Phi(y) = {\rm Im}\,\Phi(1) = -t$, and runs from the endpoint $y_1 =1$ towards the asymptotic line ${\rm arg}\,(y) = \pi - \phi$, crossing under the branch cut of multiplicity $d+1$, into the Riemann sheet lying immediately below.  The gas-saddle thimble starts at the asymptotic line ${\rm arg}\,(y) = \pi -\phi$ and runs towards the line ${\rm arg}\,(y) = -\phi$, passing through the saddle $y_g$. When merging this `near-zone' picture with the `far-zone' picture studied in the previous section, we see that in the region $y\sim O(N^{(d+1)/(2d-1)}_*)$ the 
 $\mathcal{J}_g$ contour actually gets significantly affected by the landscape of the black-hole saddles. In particular, one can see that $\mathcal{J}_g$ is attracted by the $x_b$ valley and fuses with the previous $\mathcal{J}_0$ contour, approaching asymptotically the line 
 ${\rm arg}\,(x)=-(\phi + 2\pi)/d$. Thus, the combination $\mathcal{J}_1 + \mathcal{J}_g$ is the boundary-layer resolution of the previous contour $\mathcal{J}_0$ at small values of $t$.

 The relative dominance between the endpoint at $y_1 =1$ and the gas saddle is equilibrated when ${\rm Re}\,\Phi(1) = {\rm Re}\,\Phi(y_g)$. In other words,
 \begin{equation}\label{balgas}
     -\beta + s_g \approx s_g = {s_g \over d+1} \left({\gamma\,s_g \over |\beta +it|} \right)^d \; \cos(d\phi)\;.
 \end{equation}
 Therefore, as long as $t \lesssim O(\beta)$ we can approximate $|\beta +it| \ll 1$ and the gas saddle loses dominance at angles of order
 $\phi^{(g)}_D \approx \pi / 2d \;({\rm mod}\;\pi/d)$, up to corrections of order $O\left(|\beta +it|^d\right)$. 
 
 At still higher values of $t$ we  can find the Stokes line that completely disconnects the gas saddle by solving ${\rm Im}\,\Phi(1) = {\rm Im}\,\Phi(y_g)$, or
 \begin{equation}\label{stokesg}
     t = {s_g \over d+1} \left({\gamma\,s_g \over |\beta +it|} \right)^d \; \sin(d\phi)\;.
 \end{equation}
 Again, so long as $t\approx O(\beta) \ll 1$, we can approximate the left-hand side of this equation by $t\approx 0$ to obtain $\phi^{(g)}_S \equiv\pi /d \;({\rm mod}\;\pi/d)$, up to corrections $O\left(|\beta+it|^{d+1}\right)$. After the gas saddle disconnects, the topological class of the integration contour becomes  ${\cal J}_0 + {\cal J}_B$ and,  from this point on, the topological structure of the steepest-descent contours is controlled by the black-hole landscape. 

We conclude that the gas saddle is never the dominant contribution to the full integral, losing its relevance in favor of the endpoint
even before the large AdS black-hole saddle becomes subdominant.

A standard endpoint approximation of the original integral (\ref{intp}) yields, for large $t$
 \begin{equation}\label{endpoint}
     Z(\beta+it)_{\rm endpoint} \approx {{\gamma \,s_g} \,e^{s_g -\beta-it} \over \beta+it} + O((\beta+it)^{-2})\;,
 \end{equation}
displaying power-like decay after the endpoint dominates the integral in \cref{intp}.

\section{Conclusions}
\noindent
We have analyzed the holomorphic piece of the spectral form factor, $Z(\beta+it)$, 
for conformal field theories admitting a gravity dual and defined  on ${\bf S}^{d-1}$ spheres at high temperatures. The semiclassical approximation for this quantity should give a prediction for the slope of the spectral form factor. We began by noticing that the naive analytic continuation of the large black-hole saddle value $\exp(-I_B)$ has unphysical features, particularly in $d\equiv5\; ({\rm mod}\;4)$. Next, we generalized the semiclassical approximation to the `minisuperspace' form:
\begin{equation}\label{partf}
    Z(\beta+it)\approx  \int_{x_0}^\infty d\mu(x)\, e^{-(\beta+it) M(x) + S (x)} \;,
\end{equation}
which integrates over the energy spectrum with physically motivated energy and entropy functions. In practice, this amounts to a two-component model of the dynamics: a graviton gas and small or large black holes in AdS$_{d+1}$. Our analysis of (\ref{partf}) gives the following structure for $d>3$:
\begin{itemize}
    \item For $0\leq t<t^{(g)}_D= \beta \,\tan(\pi /2d)+O({\beta^{d+1}/\ell^d})$ the thimble decomposition is given by $\mathcal{J}_1 + \mathcal{J}_g + \mathcal{J}_B$ and the partition function is dominated by the $x_B$ saddle $Z(\beta+it) \approx \exp(-I_B)$.
    \item For $t^{(g)}_D < t < t^{(B)}_D$, the gas saddle becomes subdominant with respect to the endpoint, but  the thimble structure does not change and the full integral remains dominated by the $x_B$ saddle. 
    \item For $t^{(B)}_D < t < t^{(g)}_S = \beta\,\tan (\pi/d)+O({\beta^{d+2}/\ell^{d+1}})$, the large black hole saddle becomes subdominant with respect to the $y_1 =1$ endpoint. So the full partition function is given by $Z(\beta+it) \approx Z(\beta+it)_{\rm endpoint}$ as in (\ref{endpoint}), while the thimble decomposition remains the same.  
    \item For $t^{(g)}_S < t < t^{(B)}_S = \beta\,\tan (\pi/(d-1))+O({\beta^3/\ell^2})$ the gas saddle $y_g$ has disconnected from the thimble decomposition, which becomes $\mathcal{J}_0  + \mathcal{J}_B$. The partition function remains endpoint dominated.
    \item For $t^{(B)}_S < t$, the large black-hole saddle disconnects from the thimble sum, which is just given by $\mathcal{J}'_0$. 
\end{itemize}
For the $d=3$ case, the angle of disconnection is close to $\pi/2$, which means that $t_S$ cannot be of order $\beta.$ In this case, the timescale of disconnection is provided by the radius of curvature $\ell$. The dominance, which has to do with the real part of the actions, still admits a perturbative treatment in the high-temperature regime. The endpoint structure resolution will follow that of the $d>3$ case. However, for the black hole thimbles we observe a difference. 
\begin{itemize}
    \item For $0\leq t<t^{(B)}_D= {\beta \tan\left(\frac{\pi}{4}\right)}+O(\beta^3/\ell^2)$ the thimble decomposition is given by $\mathcal{J}_0 + \mathcal{J}_B$. The partition function is dominated by the $x_B$ saddle $Z(\beta+it) \approx \exp(-I_B)$.
    \item For $t^{(B)}_D\leq t<t^{(B)}_S=  2 \pi \ell\left(1+O({\beta^2/\ell^2})\right)$ the thimble decomposition is still given by $\mathcal{J}_0 + \mathcal{J}_B$, but the partition function is endpoint dominated. 
    \item For $t>t^{(B)}_S$, the thimble topology changes: $\mathcal{J}_0\to\mathcal{J}'_0=\mathcal{J}_0+\mathcal{J}_B$, meaning the large black hole saddle disconnects. 
    
\end{itemize}

These results resolve the tension introduced by the bad behavior of $\exp(-I_B)$ under analytic continuation of the inverse temperature, as well as the potential difficulties in defining the inverse Laplace transform to obtain densities of states. In fact, once we write the partition function in the form (\ref{partf}), the inverse Laplace transform can be formally defined by  exchanging the orders of integration between $x$ and $t$, resulting in the standard form $\Omega(E) = e^{S(E)} dS/dE$. This means that, in a microcanonical analysis such as that of \cite{Barbon_2025}, one can simply generalize each saddle constructed from black-hole metrics into the family of conical non-saddles and recover the local saddle-point approximation results without worrying about the global convergence of inverse Laplace transforms. 

It would be interesting to generalize the results of this paper to other types of black holes such as AdS black holes with non-spherical horizons, stringy black holes, and also black holes with non-AdS asymptotics. 

Finally, one important topic for further research is the foundation of the starting ansatz (\ref{partf}) within the general theory of gravitational path integrals, perhaps along the lines of \cite{Marolf}. Of special relevance is the interplay between the particular contours singled out by this approximation and the general consistency conditions recently discussed in \cite{KS, W3}.\footnote{On this matter, it is intriguing that reference \cite{Chen} has identified what for us is the leading dominance angle $\phi_D = \frac{\pi}{2(d-1)}$, as the boundary of allowability of black hole near-horizon metrics in the sense of \cite{KS, W3}. }

\newpage
\section*{Acknowledgements}
\noindent
We are indebted to M. Sasieta for useful discussions and to Y. Chen for discussions and for drawing our attention to reference \cite{Chen}. 
This work is partially supported by the Severo Ochoa Program for Centers of Excellence through the grant CEX2020-001007-S and the  R\&D\&I Project CEX2025-001574-S, funded by MICIU/AEI/10.13039/ 501100011033 within the research line ``Strings and Quantum Gravity". The work was also partially supported by the grants PID2022-137127NB-I00 and PID2024-156043NB-I00, funded by MCIN/AEI/10.13039/501100011033/ FEDER, UE, and `ERDF A way of making Europe'. E. Velasco-Aja also acknowledges support from the European Union’s Horizon 2020 research and innovation programme under the Marie Sklodowska-Curie grant agreement No 860881-HIDDeN.


\end{document}